\documentclass[conference]{IEEEtran}
\IEEEoverridecommandlockouts

\usepackage{cite}
\usepackage{amsmath,amssymb,amsfonts}
\usepackage{graphicx}
\usepackage{booktabs}
\usepackage{float}
\usepackage{xcolor}
\usepackage[hidelinks]{hyperref}
\usepackage{textcomp}

\usepackage{booktabs}
\usepackage{array}
\usepackage{makecell}
\usepackage{amssymb}

\begin{document}

\title{Entangle: Uncovering Collaboration in the GitHub Quantum Software Ecosystem}

\author{
\IEEEauthorblockN{1\textsuperscript{st} \'Angel Luis Lara-Mart\'in}
\IEEEauthorblockA{\textit{Universidad de Castilla-La Mancha} \\
Talavera de la Reina, Spain \\
angelluis.lara@alu.uclm.es}
\and
\IEEEauthorblockN{2\textsuperscript{nd} Ricardo P\'erez-Castillo}
\IEEEauthorblockA{\textit{Universidad de Castilla-La Mancha} \\
Talavera de la Reina, Spain \\
ricardo.pdelcastillo@uclm.es}
}

\maketitle

\begin{abstract}
Quantum computing is moving from research laboratories towards early
commercialization and broader socio-technical adoption, supported by
sustained hardware progress and a rapidly expanding open-source software
ecosystem. This momentum is especially visible on GitHub, where many
quantum and hybrid software projects coexist around frameworks such as
Qiskit, Cirq, PennyLane and Amazon Braket. However, this ecosystem
remains fragmented, making it difficult to understand who shapes quantum
software, where expertise is concentrated, how collaboration flows across
organizations and disciplines, and which actors connect otherwise
separated communities. This paper presents Entangle, a data-driven
analysis of the open-source quantum computing ecosystem on GitHub.
Starting from 71 domain keywords, Entangle identifies more than 1{,}500
quantum repositories, 27{,}000 contributors and 400 organizations,
revealing an ecosystem strongly organized around four leading industrial
vendors, but also supported by 2{,}387 contributors who connect projects,
organizations and domains. These findings provide practical evidence for
responsible quantum innovation by making visible patterns of influence,
dependency, collaboration and knowledge transfer. They also offer
actionable indicators for strategic decisions on investment, hiring,
partnerships, ecosystem stewardship and capacity building. More broadly,
Entangle shows how open-source intelligence can support a more
transparent, measurable and governable quantum software ecosystem,
helping align technical development with responsible innovation,
public--private coordination and long-term sustainability.
\end{abstract}


\section{Introduction}
\label{sec:introduction}
Quantum computing is transitioning from research laboratories towards
early commercialization and broader socio-technical adoption. In the
Noisy Intermediate-Scale Quantum (NISQ) era~\cite{preskill2018quantum},
real progress no longer depends only on hardware: it increasingly hinges
on a thriving open-source software ecosystem built around frameworks such
as Qiskit~\cite{qiskit2023}, Cirq~\cite{cirq2023},
PennyLane~\cite{pennylane2023} and Amazon Braket~\cite{braket2023}. The
clearest expression of this momentum is GitHub, where thousands of
quantum and hybrid software projects, contributors and organizations
coexist and evolve daily.

Despite its strategic importance, this ecosystem remains fragmented and poorly understood as a whole, even if some preliminary work have attempted to research on this \cite{Umbrello24}. Activity is dispersed across thousands of loosely connected repositories, and the relationships between developers, organizations and technical domains stay largely invisible. As a result, basic but consequential questions are hard to answer: who shapes quantum software, where expertise is concentrated, how collaboration flows across organizations and disciplines, and which actors connect otherwise separated communities. Without this visibility, investment, hiring, partnership and policy decisions rely on intuition rather than evidence, and structural fragilities, such as critical projects sustained by a single maintainer, go unnoticed. From a responsible-quantum perspective, this lack of visibility is not merely an empirical gap but a governance problem: it limits transparency, accountability and ecosystem stewardship at precisely the moment when quantum technologies are moving towards commercialization and broader societal adoption \cite{Holter2023,Vishwakarma2024}.

This paper makes two contributions. First, it introduces \texttt{Entangle}, a tool-supported approach for evidence-based repository analytics in the public quantum software ecosystem on GitHub. \texttt{Entangle} turns scattered repository activity into measurable indicators of influence, dependency, collaboration and knowledge transfer by integrating repositories, contributors and organizations into a single collaboration graph. Second, the paper reports a preliminary empirical analysis conducted with \texttt{Entangle}. Starting from 71 domain keywords, the tool identifies and integrates more than 1{,}500 quantum-related repositories, 27{,}000 contributors and 400 organizations, revealing an ecosystem strongly concentrated around four industrial vendors yet connected by 2{,}387 cross-cutting contributors. We frame these indicators as decision support for responsible-quantum actors: they can prioritize maintainer support, cross-community coordination and capacity-building interventions, and subsequently monitor whether concentration and fragility indicators improve.

\section{Background and Related Work}
\label{sec:related}
Mining software repositories at ecosystem scale has a long tradition.
GHTorrent~\cite{ghtorrent2013} and GH Archive~\cite{gharchive2015} mirror
GitHub's public activity for reproducible empirical studies, but they
provide raw data rather than analysis or exploration. Dependency-centric
services such as Libraries.io~\cite{librariesio2024} and Open Source
Insights~\cite{depsdev2024} map relationships \emph{between artefacts},
not between people, and therefore do not capture collaboration dynamics.
Methodological work from the MSR community is closer in spirit:
Kalliamvakou et~al.~\cite{kalliamvakou2014github} document pitfalls of
mining GitHub, and Cohen and Consens~\cite{cohen2018cocommit} study
co-commit collaboration patterns; both inform our pipeline but stand as
isolated studies rather than ecosystem observatories. In the quantum
domain specifically, Umbrello et~al.~\cite{Umbrello24} map quantum-technology stakeholders and their interactions, but do not analyse repository-based
developer collaboration. Tavassoli Sabzevari and Khan~\cite{tavassoli2026quantum}
mine Stack Overflow to surface developer pain points. \texttt{Entangle} is
complementary: it mines development activity directly on GitHub to build
a contributor collaboration graph and turn it into governance-relevant
indicators, focusing on the \emph{social} structure of quantum software
rather than forum questions or package dependencies.
Table~\ref{tab:related} summarises this positioning: prior tools each
cover specific aspects (data collection, dependencies, security or
isolated collaboration studies), whereas \texttt{Entangle} integrates ecosystem data
ingestion, social-network analysis and interactive exploration in a single
observatory suite.

\begin{table}[!ht]
  \centering
  \caption{Entangle vs. representative ecosystem-mining tools/studies ($\checkmark$ full, $\sim$ partial, $\times$ none).}
  \label{tab:related}

  \scriptsize
  \setlength{\tabcolsep}{1.5pt}
  \renewcommand{\arraystretch}{1.15}

  \begin{tabular}{
    @{}
    >{\raggedright\arraybackslash}p{0.30\columnwidth}
    *{5}{>{\centering\arraybackslash}p{0.115\columnwidth}}
    @{}
  }
    \toprule
    \textbf{Work}
      & \textbf{Data pipeline}
      & \textbf{Social network}
      & \textbf{Dash-board}
      & \textbf{3D visualization}
      & \textbf{AI assistant} \\

    GHTorrent~\cite{ghtorrent2013}
      & $\checkmark$ & $\times$ & $\times$ & $\times$ & $\times$ \\

    GH Archive~\cite{gharchive2015}
      & $\checkmark$ & $\times$ & $\times$ & $\times$ & $\times$ \\

    Libraries.io~\cite{librariesio2024}
      & $\checkmark$ & $\times$ & $\sim$ & $\times$ & $\times$ \\

    Open Source Insights~\cite{depsdev2024}
      & $\checkmark$ & $\times$ & $\sim$ & $\times$ & $\times$ \\

    MSR studies~\cite{kalliamvakou2014github,cohen2018cocommit}
      & $\sim$ & $\checkmark$ & $\times$ & $\times$ & $\times$ \\

    Tavassoli \& Khan~\cite{tavassoli2026quantum}
      & $\checkmark$ & $\times$ & $\times$ & $\times$ & $\times$ \\

    \textbf{Entangle}
      & $\checkmark$ & $\checkmark$ & $\checkmark$
      & $\checkmark$ & $\checkmark$ \\

    \bottomrule
  \end{tabular}
\end{table}

\section{Entangle Suite}
\label{sec:entangle}
\texttt{Entangle}\footnote{\texttt{Entangle} is open-source: the
\texttt{Entangle-Core} backend
(\url{https://github.com/Angel-TFG-UCLM/Entangle-Core}) and the
\texttt{Entangle-Visualizer} frontend
(\url{https://github.com/Angel-TFG-UCLM/Entangle-Visualizer}).} operates as an end-to-end observatory whose data layer runs in six sequential, modular phases over GitHub's GraphQL
API~\cite{githubgraphql2017}. 


\begin{itemize}
\item \emph{Repository ingestion} seeds discovery with 71 quantum keywords
(e.g.\ \texttt{qiskit}, \texttt{cirq}, \texttt{openqasm}, \texttt{qubit},
\texttt{quantum machine learning}), segmenting queries by star range and
creation year to bypass the 1{,}000-result cap of the Search API.

\item \emph{Repository enrichment} gathers languages, topics, contributors, 
releases and activity.

\item \emph{User and organization ingestion} derives both entities from the 
enriched repositories.

\item \emph{Enrichment} computes a \texttt{quantum\_expertise\_score} per user 
and a \texttt{quantum\_focus\_score} per organization, and classifies each 
contributor into one of six disciplines by combining five signals: bio, company, 
organizations, repository topics and languages.

\item \emph{Network and community analysis} builds a heterogeneous collaboration 
graph with NetworkX and computes Louvain communities~\cite{louvain_method}, 
betweenness centrality~\cite{betweenness_centrality}, \textit{bridge} contributors and per-repository bus factor~\cite{avelino2016novel}.

\item \emph{Metrics caching} precomputes dashboard statistics for fast queries.

\end{itemize}

\begin{figure}[!ht]
  \centering
  \includegraphics[width=\columnwidth]{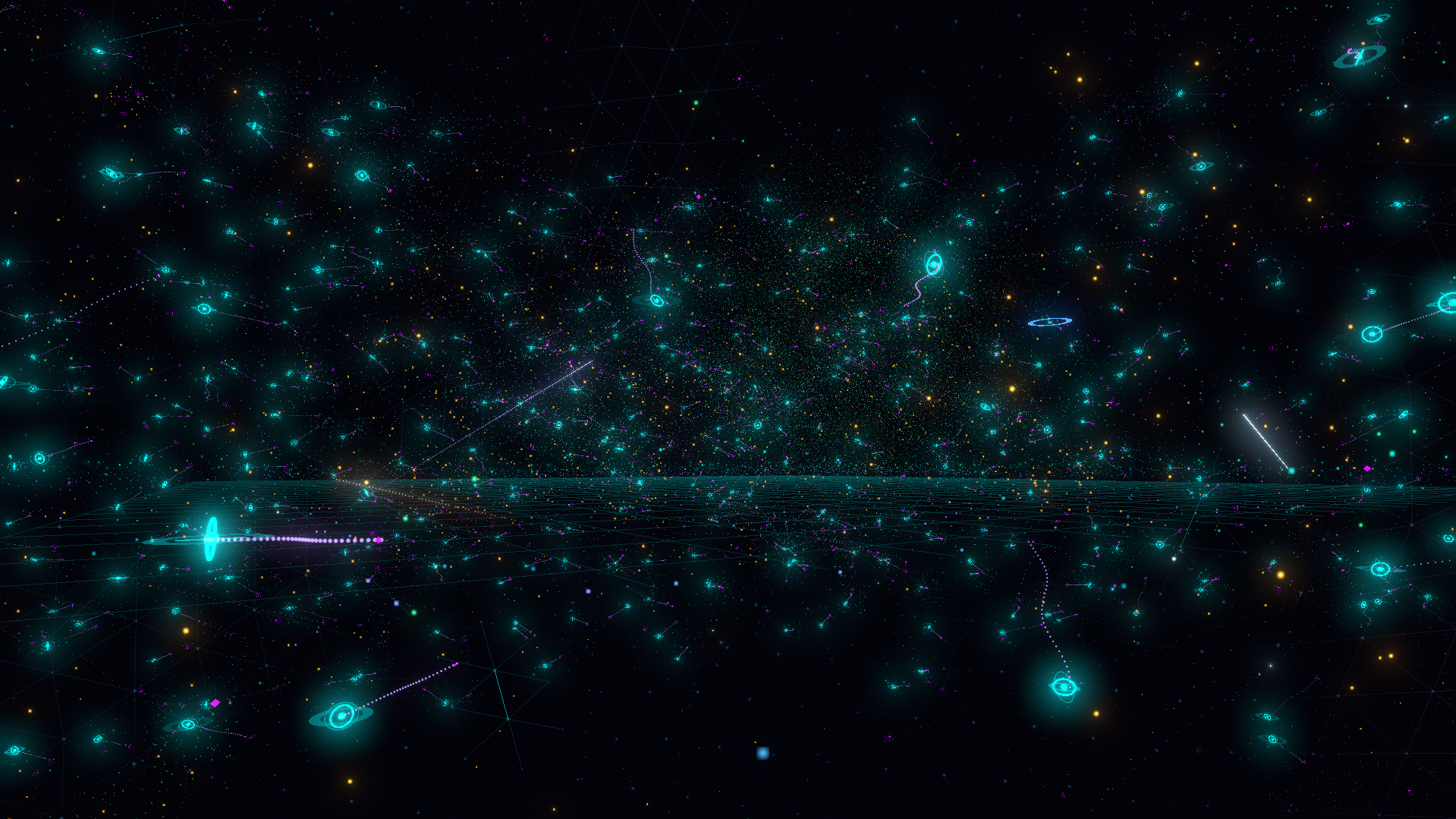}
  \caption{\texttt{Entangle}'s immersive 3D universe: the $\sim$28{,}000 ecosystem
  entities (organizations, repositories and contributors) distributed in
  space, conveying the scale and density of the open-source quantum
  ecosystem.}
  \label{fig:universe}
\end{figure}

Because the seed keywords are generic, ingestion inevitably captures false
positives (e.g.\ Firefox Quantum, games, CSS libraries); heuristic filters
over topics, description and README mitigate them. To keep the bridge
analysis meaningful, sibling organizations (e.g.\ \texttt{qiskit} and
\texttt{qiskit-community}) are identified so that contributors shared only
between them are not counted as \textit{bridges}, without merging their data; and
bot accounts are detected and flagged, which keeps them out of the bridge
computation while letting users show or hide them on demand in the
visualization. \texttt{Entangle} distinguishes two complementary notions of \emph{bridge}: \emph{bridge contributors}, active across two or more independent organizations, which capture inter-organization links; and \emph{disciplinary bridges}, contributors spanning two or more technical disciplines, which capture
cross-domain knowledge transfer. Results are exposed through three
complementary front-ends: an interactive 2D dashboard with cross-filtering,
an immersive 3D universe that distributes the $\sim$28{,}000 ecosystem
entities in space (see Fig.~\ref{fig:universe}), and a conversational agent
that queries the data in natural language. The platform is
deliberately insight-heavy: it exists to make the ecosystem
\emph{measurable}.

\section{Ecosystem Insights}
\label{sec:insights}

This section presents the main insights obtained by applying \texttt{Entangle} to public GitHub data collected in June~2026. As \texttt{Entangle} runs an incremental ingestion pipeline, this snapshot can be refreshed and extended at any time by re-running or broadening the data ingestion.

\begin{figure}[!ht]
  \centering
  \includegraphics[width=\columnwidth]{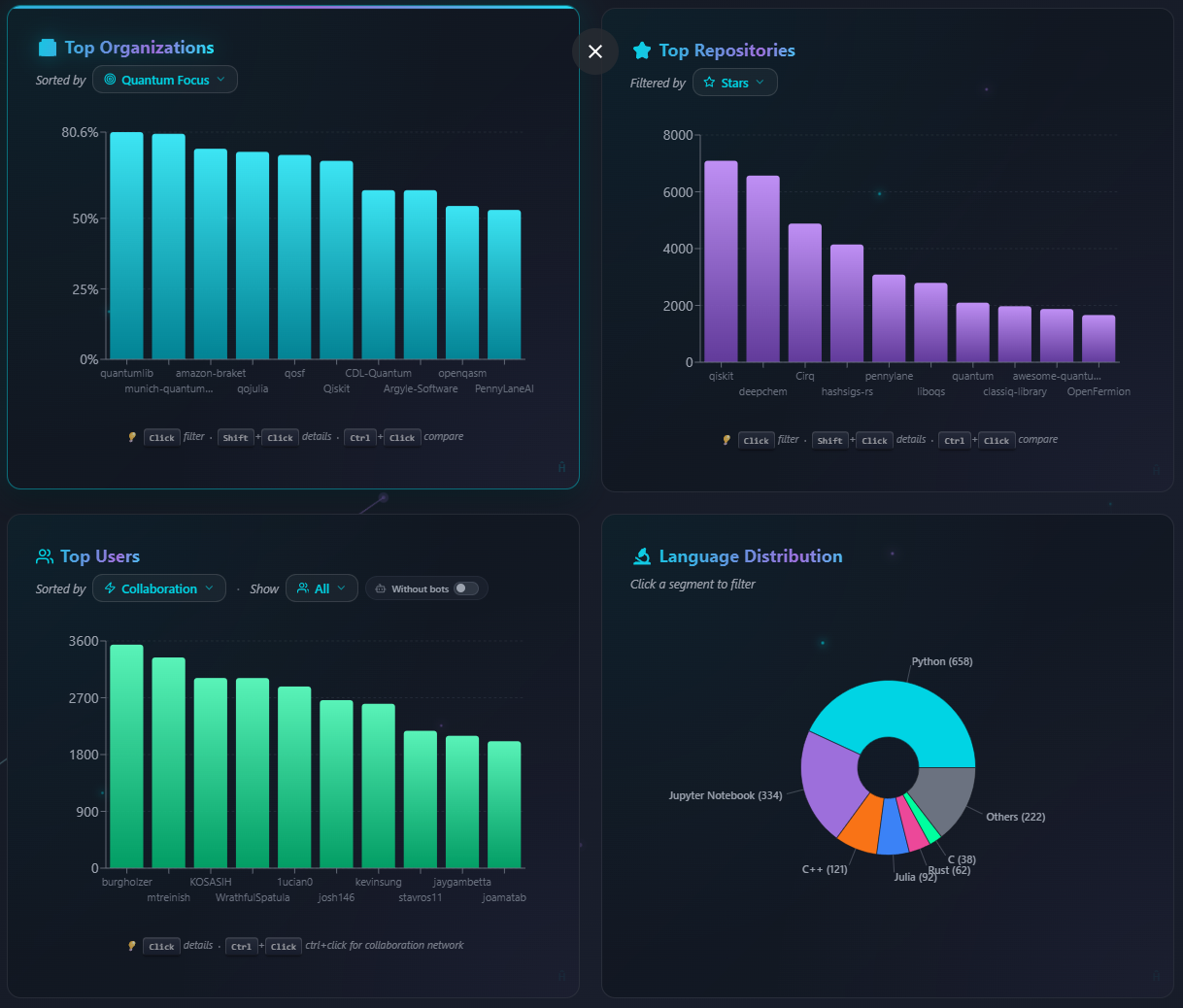}
  \caption{Analytical view: top organizations by quantum focus, top
  repositories by stars, top contributors, and language distribution,
  confirming concentration around the four leading SDK vendors.}
  \label{fig:dashboard}
\end{figure}

\begin{figure}[!ht]
  \centering
  \includegraphics[width=\columnwidth]{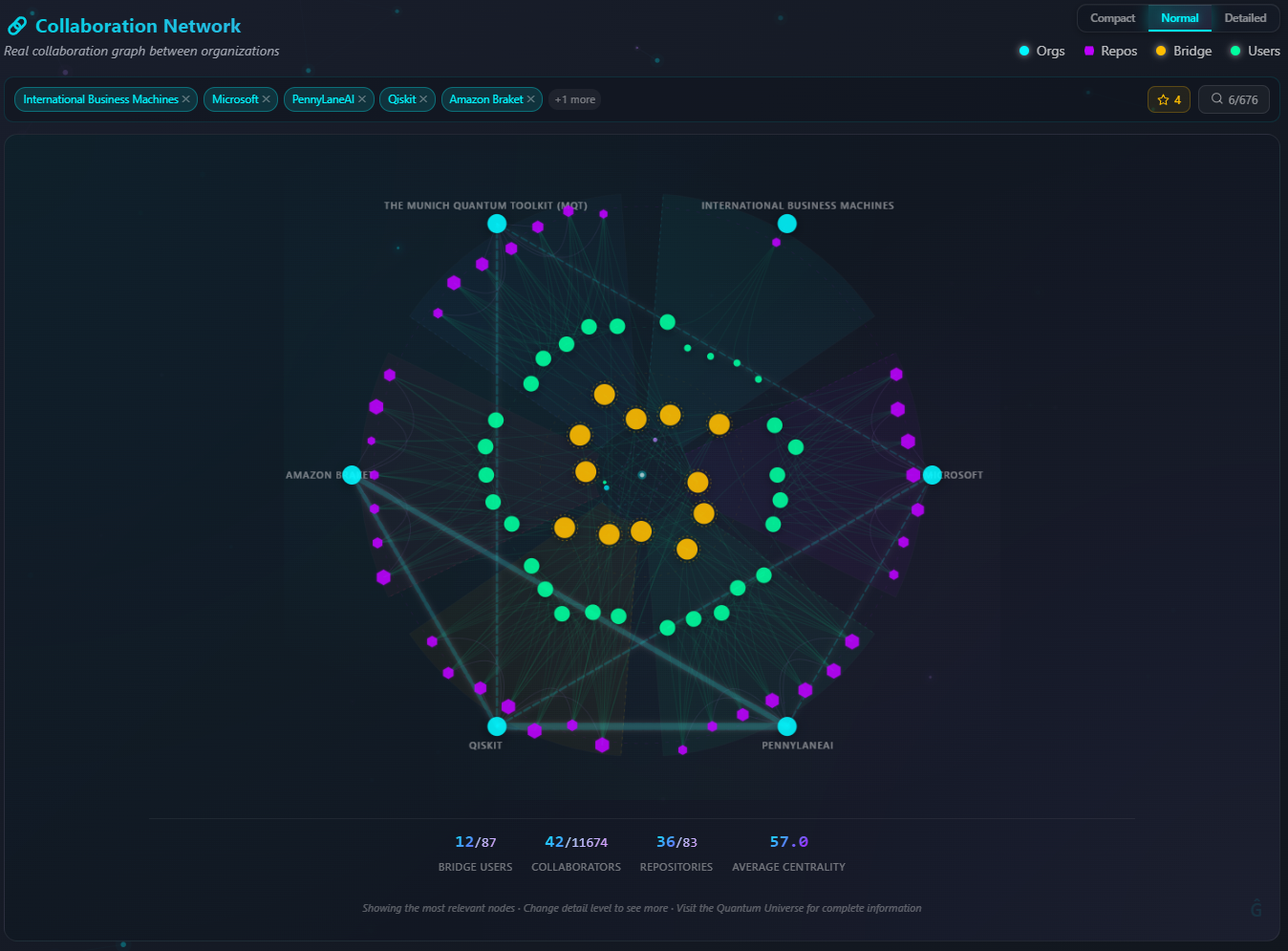}
  \caption{Collaboration mini-universe: dense organization clusters linked
  by inter-organization edges; central bridge contributors connect
  otherwise separated communities.}
  \label{fig:graph}
\end{figure}

\begin{itemize}

    \item 
\textbf{Volume and composition.} The pipeline integrates over 1{,}500
quantum repositories, 27{,}000 contributors and 400 organizations,
yielding roughly 18 contributors per repository, a sign of active but
unevenly distributed participation.
    \item 
\textbf{Languages.}
Python dominates, used as the primary language in 42.05\% of repositories
and sustained by the major SDKs; Jupyter notebooks rank second,
evidencing an experimentation- and education-driven field, while
C\texttt{++} and Julia cover high-performance simulators.

    \item 
\textbf{Vendors' Concentration.} Four vendors anchor the ecosystem: Qiskit (IBM),
Cirq (Google), PennyLane (Xanadu) and Braket (Amazon); they recur as the
highest-centrality nodes. The \texttt{quantum\_focus\_score} separates
organizations dedicated to quantum computing from those with only
tangential involvement (see Fig.~\ref{fig:dashboard}). 

    \item 
\textbf{Disciplines.} Across the six-discipline
classification, \texttt{quantum\_software} and \texttt{dev\_tooling}
jointly account for more than half of all contributors, confirming the
predominance of software development over theory, whereas
\texttt{quantum\_hardware} forms the smallest, most specialized community,
consistent with the entry barrier of physical quantum hardware.

    \item 
\textbf{Structure.} The collaboration graph
has 30{,}970 nodes and 101{,}190 edges; Louvain yields 752 communities
with modularity 0.96, i.e., tight clusters that mostly gather one
organization with its repositories and direct contributors. Two community
profiles emerge: \emph{industrial} clusters (around Qiskit/IBM,
Cirq/Google, Braket/Amazon) with frequent releases, thorough documentation
and many external collaborators; and \emph{academic} clusters with fewer
stars but more Jupyter notebooks and associated publications.

    \item
\textbf{Connectors.} A key governance-relevant finding is the existence of
2{,}387 \emph{bridge contributors}, active across two or more independent
organizations. They are not only numerous but structurally central, largely
coinciding with the highest-betweenness nodes, so they act as socio-technical
connectors across vendors, academic groups, tool-chains and application
domains. From a responsible-quantum perspective, they provide observable
evidence of where cross-organization coordination, industry--academia
knowledge transfer and ecosystem capacity building are already taking place.
The Jaccard overlap of contributors further quantifies these ties; for
instance, that between \texttt{Qiskit} and \texttt{qiskit-community} is 0.19,
i.e., 260 shared contributors (see Fig.~\ref{fig:graph}).

    \item 
\textbf{Sustainability and resilience.} The most pressing signal concerns the
ecosystem's bus factor. On a four-level scale (1, 2, 3--4, $\geq$5),
\textbf{81.7\%} of repositories have a bus factor of~1: maintenance activity
is concentrated in a single key contributor. This does not imply immediate
project failure, but reveals a systemic dependency risk if that contributor
becomes inactive. From a responsible-quantum perspective, this indicator helps
identify where maintainer support, contributor on-boarding and shared
ownership could improve the long-term resilience of public quantum software
development.

\end{itemize}

\section{Implications for Responsible Quantum Innovation}
\label{sec:discussion}
The previous results suggest that repository-level evidence can make
responsible quantum innovation more operational. Responsible innovation
in quantum computing has been framed as a practical effort to anticipate
impacts, reflect on possible concerns, engage relevant stakeholders and
influence technological trajectories before they become locked into
deployment pathways~\cite{Holter2023}. From this
perspective, \texttt{Entangle} does not merely describe the public quantum
software ecosystem on GitHub; it provides empirical signals that can inform
where responsible action may be needed. \texttt{Entangle} can be used by funders, maintainers and institutions
to prioritize concrete stewardship interventions: supporting
single-maintainer projects, strengthening weak cross-organizational links,
and developing weakly represented disciplines or communities. Its indicators
can then be recomputed to monitor whether concentration, participation
breadth and resilience improve after those interventions. This creates a
feedback loop between ecosystem evidence and responsible action,
operationalizing anticipation, stakeholder engagement and equitable capacity
building~\cite{Holter2023,Vishwakarma2024}.

Several limitations bound the interpretation of these results. Discovery
is keyword-driven: generic seeds capture false positives (e.g.\ Firefox
Quantum), mitigated but not eliminated by heuristic filtering over topics,
description and \texttt{README}, so absolute counts should be read as
well-founded estimates. Moreover, the analysis is a snapshot of a fast-moving ecosystem; longitudinal tracking is needed to confirm whether the observed structure is stable over time. \texttt{Entangle} observes only public GitHub activity, so commercially sensitive or proprietary development hosted in
private repositories may be absent, limiting the completeness of
organization- and ecosystem-level conclusions. Finally, the bus factor is a heuristic proxy for maintenance risk based on contribution concentration~\cite{avelino2016novel} rather than a direct measure of project health, and \textit{bridge} detection depends on the sibling-organization and bot-detection rules, which, although conservative, are themselves heuristics. These threats motivate the open release of reproducible indicators as future work.

\section{Conclusions and Future Work}
\label{sec:conclusions}

\texttt{Entangle} shows that the publicly visible quantum software
ecosystem on GitHub can be mapped at scale through evidence-based
repository analytics. The main contribution is the tool-supported approach
itself: \texttt{Entangle} integrates repositories, contributors and
organizations into a collaboration graph, producing indicators of
influence, dependency, collaboration and knowledge transfer. As a second
contribution, we report a preliminary empirical analysis conducted with
the tool, revealing a productive but structurally fragile ecosystem:
activity is concentrated around four major vendors, the graph is highly
modular, 2{,}387 \textit{bridge} contributors connect organizations and
domains, and 81.7\% of the analyzed repositories have a bus factor of~1.

Beyond descriptive ecosystem mapping, organizations can use
\texttt{Entangle} to examine their publicly visible repository portfolio
and collaboration network, identify single-maintainer projects, weak
cross-team connections and concentrated expertise, and address these
risks before they become critical points of failure. Recomputing the
indicators after such interventions would also allow organizations to
monitor whether resilience improves. In this sense, \texttt{Entangle}
complements principle-based discussions of responsible quantum with
measurable signals that can inform governance, community building and
sustainable capacity development in quantum software..

Future work will focus on turning this static snapshot into a living
observatory. We plan to incorporate longitudinal analyses of repository
activity, community evolution and dependency drift; improve discovery
through adaptive keyword expansion; and strengthen graph analysis with
well-connected community detection methods such as
Leiden~\cite{traag2019leiden}. We also intend to co-design these decision workflows with maintainers, funders and institutions, and evaluate whether the indicators are understandable, actionable and useful for monitoring responsible-innovation interventions.

\section*{Acknowledgments} 
This work is funded by projects Smooth (PID2022-137944NB-I00) (MICIU/AEI/10.13039/501100011033/PRTR, EU), and Sinergia (2025-GRIN-38310) (UCLM and ERDF).

\bibliographystyle{IEEEtran}
\bibliography{references}

@article{preskill2018quantum,
  author  = {Preskill, John},
  title   = {Quantum Computing in the {NISQ} Era and Beyond},
  journal = {Quantum},
  volume  = {2},
  pages   = {79},
  year    = {2018},
  doi     = {10.22331/q-2018-08-06-79}
}

@misc{qiskit2023,
  author = {{IBM Quantum}},
  title  = {Qiskit: Open-Source Quantum Development},
  year   = {2024},
  note   = {https://qiskit.org/}
}

@misc{cirq2023,
  author = {{Google Quantum AI}},
  title  = {Cirq: A Python Framework for Quantum Circuits},
  year   = {2024},
  note   = {https://quantumai.google/cirq}
}

@misc{pennylane2023,
  author = {{Xanadu}},
  title  = {PennyLane: A Cross-Platform Library for Quantum Computing},
  year   = {2024},
  note   = {https://pennylane.ai/}
}

@misc{braket2023,
  author = {{Amazon Web Services}},
  title  = {Amazon Braket: Quantum Computing Service},
  year   = {2024},
  note   = {https://aws.amazon.com/braket/}
}

@article{ghtorrent2013,
  author  = {Gousios, Georgios},
  title   = {The {GHTorrent} Dataset and Tool Suite},
  journal = {Proc. 10th Working Conf. Mining Software Repositories (MSR)},
  pages   = {233--236},
  year    = {2013},
  doi     = {10.1109/MSR.2013.6624034}
}

@misc{gharchive2015,
  author = {Grigorik, Ilya},
  title  = {{GH} Archive: Recording the Public {GitHub} Timeline},
  year   = {2015},
  note   = {https://www.gharchive.org/}
}

@misc{librariesio2024,
  author = {{Libraries.io}},
  title  = {{Libraries.io}: Open Source Discovery Service},
  year   = {2024},
  note   = {https://libraries.io/}
}

@misc{depsdev2024,
  author = {{Google}},
  title  = {Open Source Insights ({deps.dev})},
  year   = {2024},
  note   = {https://deps.dev/}
}

@article{kalliamvakou2014github,
  author  = {Kalliamvakou, Eirini and Gousios, Georgios and Blincoe, Kelly and Singer, Leif and German, Daniel M. and Damian, Daniela},
  title   = {The Promises and Perils of Mining {GitHub}},
  journal = {Proc. 11th Working Conf. Mining Software Repositories (MSR)},
  pages   = {92--101},
  year    = {2014},
  doi     = {10.1145/2597073.2597074}
}

@inproceedings{tavassoli2026quantum,
  author    = {Tavassoli Sabzevari, Maryam and Khan, Arif Ali},
  title     = {Empirical Investigation of Quantum Computing Toolchains and Algorithms: Mining Stack Overflow},
  booktitle = {Companion Proc. 34th ACM Int. Conf. Foundations of Software Engineering (FSE)},
  year      = {2026},
  doi       = {10.48550/arXiv.2604.15512}
}

@article{louvain_method,
  author  = {Blondel, Vincent D. and Guillaume, Jean-Loup and Lambiotte, Renaud and Lefebvre, Etienne},
  title   = {Fast Unfolding of Communities in Large Networks},
  journal = {J. Stat. Mech.: Theory and Experiment},
  volume  = {2008},
  number  = {10},
  pages   = {P10008},
  year    = {2008},
  doi     = {10.1088/1742-5468/2008/10/P10008}
}

@article{betweenness_centrality,
  author  = {Freeman, Linton C.},
  title   = {A Set of Measures of Centrality Based on Betweenness},
  journal = {Sociometry},
  volume  = {40},
  number  = {1},
  pages   = {35--41},
  year    = {1977},
  doi     = {10.2307/3033543}
}

@inproceedings{avelino2016novel,
  author    = {Avelino, Guilherme and Passos, Leonardo and Hora, Andre and Valente, Marco Tulio},
  title     = {A Novel Approach for Estimating Truck Factors},
  booktitle = {Proc. 24th Int. Conf. Program Comprehension (ICPC)},
  pages     = {1--10},
  year      = {2016},
  doi       = {10.1109/ICPC.2016.7503718}
}

@inproceedings{cohen2018cocommit,
  author    = {Cohen, Eldan and Consens, Mariano P.},
  title     = {Large-Scale Analysis of the Co-Commit Patterns of {GitHub}'s Top Repositories},
  booktitle = {Proc. 15th Int. Conf. Mining Software Repositories (MSR)},
  pages     = {426--436},
  year      = {2018},
  doi       = {10.1145/3196398.3196436}
}

@misc{githubgraphql2017,
  author = {{GitHub, Inc.}},
  title  = {{GitHub GraphQL API} Documentation},
  year   = {2024},
  note   = {https://docs.github.com/en/graphql}
}

@article{traag2019leiden,
  author  = {Traag, Vincent A. and Waltman, Ludo and van Eck, Nees Jan},
  title   = {From {Louvain} to {Leiden}: Guaranteeing Well-Connected Communities},
  journal = {Scientific Reports},
  volume  = {9},
  number  = {1},
  pages   = {5233},
  year    = {2019},
  doi     = {10.1038/s41598-019-41695-z}
}

@article{Holter2023,
    author = {Carolyn Ten Holter and Philip Inglesant and Marina Jirotka},
    title = {Reading the road: challenges and opportunities on the path to responsible innovation in quantum computing},
    journal = {Technology Analysis \& Strategic Management},
    volume = {35},
    number = {7},
    pages = {844--856},
    year = {2023},
    publisher = {Routledge},
    doi = {10.1080/09537325.2021.1988070},
    URL = {https://doi.org/10.1080/09537325.2021.1988070},
    eprint = {https://doi.org/10.1080/09537325.2021.1988070}
}

@INPROCEEDINGS{Vishwakarma2024,
  author={Vishwakarma, Saniav and Baine, Vishal Sharathchandra and Mandelbaum, Ryan and Kobayashi, Yuri and Lanes, Olivia and Wolf-Bauwens, Mira Luca},
  booktitle={2024 IEEE International Conference on Quantum Computing and Engineering (QCE)}, 
  title={The Role of Community Building and Education as Key Pillar of Institutionalizing Responsible Quantum}, 
  year={2024},
  volume={02},
  number={},
  pages={86-91},
  doi={10.1109/QCE60285.2024.10258}}

@article{Umbrello24,
author = {Umbrello, Steven and Seskir, Zeki C. and Vermaas, Pieter E.},
title = {Communities of quantum technologies: Stakeholder identification, legitimation and interaction},
journal = {International Journal of Quantum Information},
volume = {22},
number = {07},
pages = {2450012},
year = {2024},
doi = {10.1142/S0219749924500126},

URL = { 
    
        https://doi.org/10.1142/S0219749924500126
    
    

},
eprint = { 
    
        https://doi.org/10.1142/S0219749924500126
}
}

\end{document}